\documentclass[10pt]{emulateapj}
\usepackage{apjfonts}
\usepackage{amsmath}
\usepackage{rotating}
\usepackage{enumerate}
\usepackage{natbib}
\newcommand{\pivec}{\mbox{\boldmath $\pi$}}
\newcommand{\muvec}{\mbox{\boldmath $\mu$}}

\lefthead{} 
\righthead{}

\begin{document}

\title{
OGLE-2016-BLG-1045: A Test of Cheap Space-Based Microlens Parallaxes
}

\author{
I.-G.~Shin\altaffilmark{H1,K0},
A.~Udalski\altaffilmark{O1,O0},
J.~C.~Yee\altaffilmark{H1,K0,S0,F0},
S.~Calchi~Novati\altaffilmark{S1,S0},
G.~Christie\altaffilmark{F1,F0},\\
AND\\
R.~Poleski\altaffilmark{O1,E1},
P.~Mr\'oz\altaffilmark{O1},
J.~Skowron\altaffilmark{O1},
M.~K.~Szyma\'nski\altaffilmark{O1},
I.~Soszy\'nski\altaffilmark{O1},
P.~Pietrukowicz\altaffilmark{O1},
S.~Koz{\l}owski\altaffilmark{O1},
K.~Ulaczyk\altaffilmark{O1,O2},
M.~Pawlak\altaffilmark{O1}\\
(OGLE Collaboration),\\
T.~Natusch\altaffilmark{F2,F1},
R.~W.~Pogge\altaffilmark{E1},\\
($\mu$FUN Collaboration),\\
A.~Gould\altaffilmark{K1,E1,E2}, 
C.~Han\altaffilmark{K5},
M.~D.~Albrow\altaffilmark{K3},
S.-J.~Chung\altaffilmark{K1,K2},
K.-H.~Hwang\altaffilmark{K1},
Y.-H.~Ryu\altaffilmark{K1},
Y.~K.~Jung\altaffilmark{H1},
W.~Zhu\altaffilmark{E1},
C.-U.~Lee\altaffilmark{K1,K2},  
S.-M.~Cha\altaffilmark{K1,K4},   
D.-J.~Kim\altaffilmark{K1},
H.-W.~Kim\altaffilmark{K1,K2},
S.-L.~Kim\altaffilmark{K1,K2},
Y.~Lee\altaffilmark{K1,K4},
D.-J.~Lee\altaffilmark{K1},
B.-G.~Park\altaffilmark{K1,K2},\\
(KMTNet Group),\\
C.~Beichman\altaffilmark{S2},
G.~Bryden\altaffilmark{S3},
S.~Carey\altaffilmark{S4},
B.~S.~Gaudi\altaffilmark{E1},
C.~B.~Henderson\altaffilmark{S3,N0},
Y.~Shvartzvald\altaffilmark{S3,N0}\\
({\it Spitzer} Team)\\
}

\bigskip\bigskip
\affil{$^{H1}$Harvard-Smithsonian Center for Astrophysics, 60 Garden St., Cambridge, MA 02138, USA}

\affil{$^{O1}$Warsaw University Observatory, Al. Ujazdowskie 4, 00-478 Warszawa, Poland}
\affil{$^{O2}$Department of Physics, University of Warwick, Gibbet Hill Road, Coventry CV4 7AL, UK}

\affil{$^{F1}$Auckland Observatory, Auckland, New Zealand}
\affil{$^{F2}$Institute for Radio Astronomy and Space Research (IRASR), AUT University, Auckland, New Zealand} 

\affil{$^{K1}$Korea Astronomy and Space Science Institute, 776 Daedeokdae-ro, Yuseong-Gu, Daejeon 34055, Korea}
\affil{$^{K2}$Korea University of Science and Technology, 217 Gajeong-ro, Yuseong-gu, Daejeon 34113, Korea}
\affil{$^{K3}$University of Canterbury, Department of Physics and Astronomy, Private Bag 4800, Christchurch 8020, New Zealand}
\affil{$^{K4}$School of Space Research, Kyung Hee University, Giheung-gu, Yongin, Gyeonggi-do, 17104, Korea}
\affil{$^{K5}$Department of Physics, Chungbuk National University, Cheongju 28644, Korea}

\affil{$^{S1}$ IPAC, Mail Code 100-22, California Institute of Technology, 1200 E. California Boulevard, Pasadena, CA 91125, USA}
\affil{$^{S2}$ NASA Exoplanet Science Institute, California Institute of Technology, Pasadena, CA 91125, USA}
\affil{$^{S3}$ Jet Propulsion Laboratory, California Institute of Technology, 4800 Oak Grove Drive, Pasadena, CA 91109, USA}
\affil{$^{S4}$ Spitzer Science Center, MS 220-6, California Institute of Technology, Pasadena, CA, USA}

\affil{$^{E1}$Department of Astronomy, Ohio State University, 140 W. 18th Ave., Columbus, OH 43210, USA}
\affil{$^{E2}$Max-Planck-Institute for Astronomy, K\"onigstuhl 17, 69117 Heidelberg, Germany}

\affil{$^{O0}$OGLE Collaboration}
\affil{$^{F0}$$\mu$FUN Collaboration}
\affil{$^{K0}$KMTNet Group}
\affil{$^{S0}${\it Spitzer} Team}
\affil{$^{N0}$NASA Postdoctoral Program Fellow}

\begin{abstract}
Microlensing is a powerful and unique technique to probe isolated objects in the Galaxy. To study 
the characteristics of these interesting objects based on the microlensing method, measurement of 
the microlens parallax is required to determine the properties of the lens. Of the various methods 
to measure microlens parallax, the most routine way is to make simultaneous ground- and space-based 
observations, i.e., by measuring the space-based microlens parallax. However, space-based campaigns 
usually require ``expensive'' resources. \citet{gould12} proposed an idea called the ``cheap space-based 
microlens parallax'' that can measure the lens-parallax using only {\it two} or {\it three} space-based observations 
of high-magnification events (as seen from Earth). This cost-effective observation strategy to measure microlens parallaxes 
could be used by space-borne telescopes to build a complete sample for studying isolated objects. This 
would enable a direct measurement of the mass function including both extremely low-mass objects and 
high-mass stellar remnants. However, to adopt this idea requires a test to check how it would work in 
actual situations. Thus, we present the first practical test of this idea using the high-magnification 
microlensing event OGLE-2016-BLG-1045, for which a subset of {\it Spitzer} observations fortuitously 
duplicate the prescription of \citet{gould12}. From the test, we confirm that the measurement of the 
lens-parallax adopting this idea has sufficient accuracy to determine the physical properties of the 
isolated lens.
\end{abstract}


\section{Introduction}

 Isolated objects with various masses such as free-floating planets, brown dwarfs, and black holes 
are very interesting targets (or potential targets) of study. At the low-mass end, free-floating 
planets and brown dwarfs may represent the low-mass tail of star formation or the result of bodies 
ejected during planet formation. Larger-mass objects ($\gtrsim\,$ several Jupiter masses) have been 
found with direct imaging in star-forming regions \citep[e.g.,][]{bihain09,esplin17}, and there exist 
several scenarios to explain their origin and evolution depending on various environmental factors 
\citep{whitworth07}. Microlensing has also probed the free-floating planet population, but with 
contradictory results. \citet{sumi11} argued that Jupiter-mass free-floating planets are about 
twice as numerous as stars, but \citet{mroz17} did not find any evidence for such a population. 
At the same time, \citet{mroz17,mroz18} discovered several candidates for less massive (few Earth-mass) 
free-floating planets. These lower mass objects could be candidates for ejection from forming 
planetary systems \citep[e.g.,][]{juric08, chatterjee08, barclay17}.
 
 At the high-mass end, there is tension between theoretical predictions of the stellar remnant 
distribution and the observed population inferred from close binaries. \citet{fryer12} predict 
a smooth distribution of remnant masses ranging from neutron stars to the most massive stellar 
mass black holes. In contrast, \citet{ozel12} find a distinct gap between the neutron star and 
black hole populations in the interval from $\sim 2$ -- $5$ $M_\odot$. Because the only confirmed 
black holes are found in binary systems, it is unclear whether this feature (and this conflict 
between observation and theory) is intrinsic to the mass distribution or somehow specific 
to stellar remnants in close binaries.

 Observations of isolated objects spanning the full mass function are necessary to resolve these 
issues. Despite the interest of these objects, their discovery and study are challenging because 
they are generally too faint to find (or they may be entirely dark). Moreover, they have no interaction 
with other stellar objects. Compared to other methods, the microlensing technique is a powerful 
and unique tool to probe these isolated objects because the technique can in principle detect 
any object that approaches or aligns with the line of sight between a background star (source) 
and observer(s), regardless of the brightness of the objects (lenses). 

 Unfortunately, microlensing observations do not, by themselves, routinely measure the microlens 
mass, $M$. Rather, they usually return only the Einstein timescale $t_{\rm E}$, which is a combination 
of several physical properties of the lens-source system
\begin{equation}
t_{\rm E} \equiv {\theta_{\rm E}\over\mu_{\rm rel}};~~
\theta_{\rm E} \equiv \sqrt{\kappa M \pi_{\rm rel}};~~
\kappa \equiv {4G\over c^2\,{\rm au}}\simeq 8.144\,{{\rm mas}\over M_\odot}.
\label{eqn:tedef}
\end{equation}
Here, ($\pi_{\rm rel}, \muvec_{\rm rel}$) are the lens-source relative
(parallax, proper motion) and $\mu_{\rm rel}=|\muvec_{\rm rel}|$.
Equation~(\ref{eqn:tedef}) implies that to determine the mass $M$
of dark (or at least, unseen) lenses, requires the measurement
of both the Einstein radius $\theta_{\rm E}$ and the
scalar amplitude $\pi_{\rm E}= |\pivec_{\rm E}|$ of the vector microlens parallax
\begin{equation}
\pivec_{\rm E} \equiv {\pi_{\rm rel}\over\theta_{\rm E}}\,{\muvec_{\rm rel}\over \mu_{\rm rel}};~~
M={\theta_{\rm E}\over\kappa\pi_{\rm E}};~~
\pi_{\rm rel} = \theta_{\rm E}\pi_{\rm E}.
\label{eqn:piedef}
\end{equation}

 According to Equation~(\ref{eqn:piedef}), the microlens parallax quantifies the lens-source vector 
displacement as seen from different observers' positions, relative to the size of the angular Einstein 
ring radius. The displacements can be caused by the annual motion of Earth, i.e., the annual microlens 
parallax \citep[hereafter APRX;][]{gould92}, different locations of observatories, such as Earth compared to 
space-borne telescopes, i.e., the space-based microlens parallax \citep[hereafter SPRX;][]{refsdal66}, or different 
ground-based sites, i.e., the terrestrial microlens parallax \citep[hereafter TPRX;][]{gould97}.

 Each method to measure microlens parallaxes has its limitations.
The APRX method \citep{alcock95,mao99,smith02} requires enough time for the motion of Earth 
to displace the observer's position from rectilinear motion enough to measure the parallax. 
As a result, the APRX can be measured for long timescale events with timescales 
$t_{\rm E}\ga30\,$days in favorable cases, but usually $t_{\rm E}\ga60\,$days. However, 
these long timescale events are not common. Moreover, from Equations~(\ref{eqn:tedef}) and 
(\ref{eqn:piedef}), this method can almost never be applied to low-mass lenses. 
For the TPRX, the displacement can be provided by a combination of simultaneous observations 
from ground-based telescopes that are well separated. However, because the size of Earth is 
only a tiny fraction of the projected Einstein ring on the observer plane 
($R_{\oplus} \ll \tilde{r}_{\rm E} \equiv {\rm au}/\pi_{\rm E}$), this measurement can be made 
for only a few special cases, i.e., extremely magnified lensing events \citep{gould09}, 
for which the strongly divergent magnification pattern is very sensitive to small changes 
in position. Thus, unfortunately, the chance for TPRX measurements would be extremely rare 
\citep{gould13}. 

 The SPRX method can provide a ``routine opportunity'' for measuring the microlens parallax as 
compared to the low chance of measuring lens-parallax with the other methods of the lens-parallax 
measurements (APRX and TPRX). This is because the displacement of the space-based observatory from 
the Earth can easily be a significant fraction of the Einstein ring, e.g., {\it Spitzer} is 
$\sim 1.3$ au from Earth compared to a typical value of $\tilde{r}_{\rm E} \sim 10$ au.
\citet{refsdal66} already proposed this method a half century ago, and \citet{dong07} made 
the first such measurement. Beginning in $2014$, the {\it Spitzer} satellite has observed 
more than $500$ microlensing events with this aim, yielding almost $80$ published microlens 
parallaxes \citep{bozza16, calchi15a, chung17, han16, han17, poleski16, ryu18, shin17, 
shvartzvald15, shvartzvald16, shvartzvald17, street16, udalski15b, wang17, yee15a, zhu15, 
zhu16, zhu17}. Even though the SPRX can provide a robust opportunity for measuring microlens 
parallaxes, there still remains an obstacle to regular adoption of the method because space-based 
observations usually require ``expensive'' resources.

 \citet{gould12} (hereafter, GY12) proposed to measure ``cheap space-based microlens parallaxes 
(cheap-SPRX)'' for high-magnification events (as seen from Earth). They showed that because the 
lens-source separation (scaled to $\theta_{\rm E}$) $u$ is extremely small near the peak of 
a high-magnification $A_{\rm max}\gg 1$ event, $u_{0,\oplus}\simeq A^{-1}\rightarrow 0$, the 
magnitude of the SPRX ($\pi_{\rm E}$) is given by 
\begin{equation}
\pi_{\rm E} \simeq {{\rm au} \over D_{\rm sat}}u_{\rm sat}
\end{equation}
\begin{equation}
u_{\rm sat} = \sqrt{2[(1-A_{\rm sat}^{-2})^{-1/2}-1]} \sim A_{\rm sat}^{-1}.
\end{equation}
Here, $D_{\rm sat}$ is the known projected (on the plane of the sky) separation to the satellite, 
e.g., $D_{\rm sat} \simeq 1.3$ au for the {\it Spitzer} space telescope, and $u_{\rm sat}$ is the 
position of satellite in the Einstein ring at the exact moment of the peak of the event as seen 
from Earth. Space-based observations can be used to determine $u_{\rm sat}$ based on $A_{\rm sat}$,
\begin{equation}
A_{\rm sat} = {{F_{\rm sat} - F_{\rm base,sat}} \over F_{\rm s,sat}} + 1.
\end{equation}
The space-based observations provide the $F_{\rm sat}$ (from an observation at the ground-based peak) 
and $F_{\rm base,sat}$ (from an observation at ``baseline'', i.e., well after the event), and 
ground-based observations can be used to constrain the source $F_{\rm s,sat}$ through color-constraints 
\citep{calchi15b,gould10a}. Hence, we can efficiently determine the magnitude of the microlens parallax 
for high-magnification events.

 The cheap-SPRX is ``cheap'' in two senses. First, as described in GY12, only two or three space-based 
observed data points are required to measure the microlens parallax. Second, this technique can be applied 
to only a small fraction of events \citep[the total number of high-magnification events is inversely proportional 
to the peak magnification;][]{gould10b}. Hence, if a satellite in solar orbit could be equipped with a camera and a means for 
prompt response for observations, it could carry out such a program at tiny additional cost to its principal mission.

 GY12 discussed a potential application of the cheap-SPRX: to study planets through the high-magnification 
channel. High-magnification events are required for the cheap-SPRX, and they are a very important channel 
to discover planets because this channel provides almost $100$ per cent detection efficiency if the events 
contain planetary mass companions to the lens stars \citep{griest98}. Based on these findings, GY12 argued 
that the cheap-SPRX could yield an unbiased measurement of the distribution of planets in the Galaxy.

 However, since that time, a second major application has emerged: the mass function of isolated objects 
in the Galaxy (particularly, for low-mass objects). The masses of isolated objects can be measured only 
if the finite source effect is observed, i.e., if $u_0 \la \rho_{\ast}$, where 
$\rho_{\ast}\equiv\theta_{\ast}/ \theta_{\rm E}$ and $\theta_{\ast}$ is the angular radius of the source. 
This generally requires a high-magnification event (since $\rho_{\ast}$ is typically 
$\mathcal{O}(10^{-3}$ -- $10^{-2})$. This is the same condition necessary to measure the cheap-SPRX. 
\citet{gould97} had already noted that high-magnification events could be used to yield isolated masses 
from a combination of finite source effects and the TPRX. Moreover, two cases were actually observed 
\citep{gould09,yee09}. \citet{gould13} showed the number of these measurements should be $\propto n$, 
where $n$ is the number density of objects, compared to the underlying microlensing event rate 
$\propto n\sqrt{M}$, where $M$ is the lens mass. Hence, they are especially useful for measuring the mass 
function of low-mass objects because these are the most abundant objects in the Galaxy. However, 
as mentioned above, the chance of measuring such a TPRX is extremely low. Thus, in a practical sense, 
the study of isolated objects cannot be effectively carried out using the TPRX alone.

 Compared to measurements of the TPRX, the SPRX can provide more robust opportunities to make the measurements. 
Actually, using {\it Spitzer} observations, \citet{zhu16} and \citet{chung17} found that a remarkably high 
fraction $(3/170)$ of $2015$ {\it Spitzer} targets yielded such isolated mass measurements. The principal reason 
is that {\it Spitzer} enables parallax measurements of much larger sources. For TPRX, by contrast, \citet{gould13} 
showed that the maximum lens distance for which the method could be applied for large sources scales as 
$D_{\rm L}\propto \theta_{\ast}^{-1}$, implying that the available volume scales as $\theta_{\ast}^{-3}$, thus 
virtually eliminating large sources for this method. These larger sources have a higher cross-section for crossing 
the lens, so a better chance of observing finite source effects\footnote[1]{\citet{zhu16} also noted that for standard 
SPRX, it is also more likely to see the finite source effect because there are two different observatory positions. 
However, this advantage is not relevant to cheap-SPRX.}.

 In fact, {\it Spitzer} itself is not well matched to the task of systematically measuring cheap-SPRX 
for high magnification events. {\it Spitzer} observations require long lead times ($3-10$ day delay 
between target selections and start of those observations, see Figure~1 of \citealt{udalski15b}), 
which raises the possibility of missing very short timescale events, which are most likely to be 
caused by the lowest mass objects. Moreover, {\it Spitzer} can observe the bulge only six weeks 
out of the eight month bulge season. In addition, the final campaign is currently scheduled 
to be in $2018$.

 As mentioned above, a systematic campaign to measure the cheap-SPRX could be conducted as an ``add-on'' 
capability to some future space mission. This would greatly increase the fraction of isolated objects 
characterized by microlensing. Based on this sample, we can determine the mass function of isolated objects 
at low cost. However, before pursuing such a course, we should perform a practical test of the cheap-SPRX 
idea to check the accuracy of the microlens parallax measurement. This test is important because 
the accuracy that can be achieved is directly related to establishing the feasibility of applying the cheap-SPRX 
under actual conditions and also for establishing an observational strategy for such a future, 
space-based microlensing campaign.
 
 Here, we conduct the first practical test for the cheap-SPRX idea using the microlensing event OGLE-2016-BLG-1045 
with {\it Spitzer} observations. In Section 2.1, we describe the event as a testbed for this practical test. 
In Section 2.2, we describe our method for testing the idea. Then, we present test results and our findings in 
Section 2.3. Lastly, we conclude and discuss in Section 3.

\begin{figure*}[htb!]
\epsscale{0.90}
\plotone{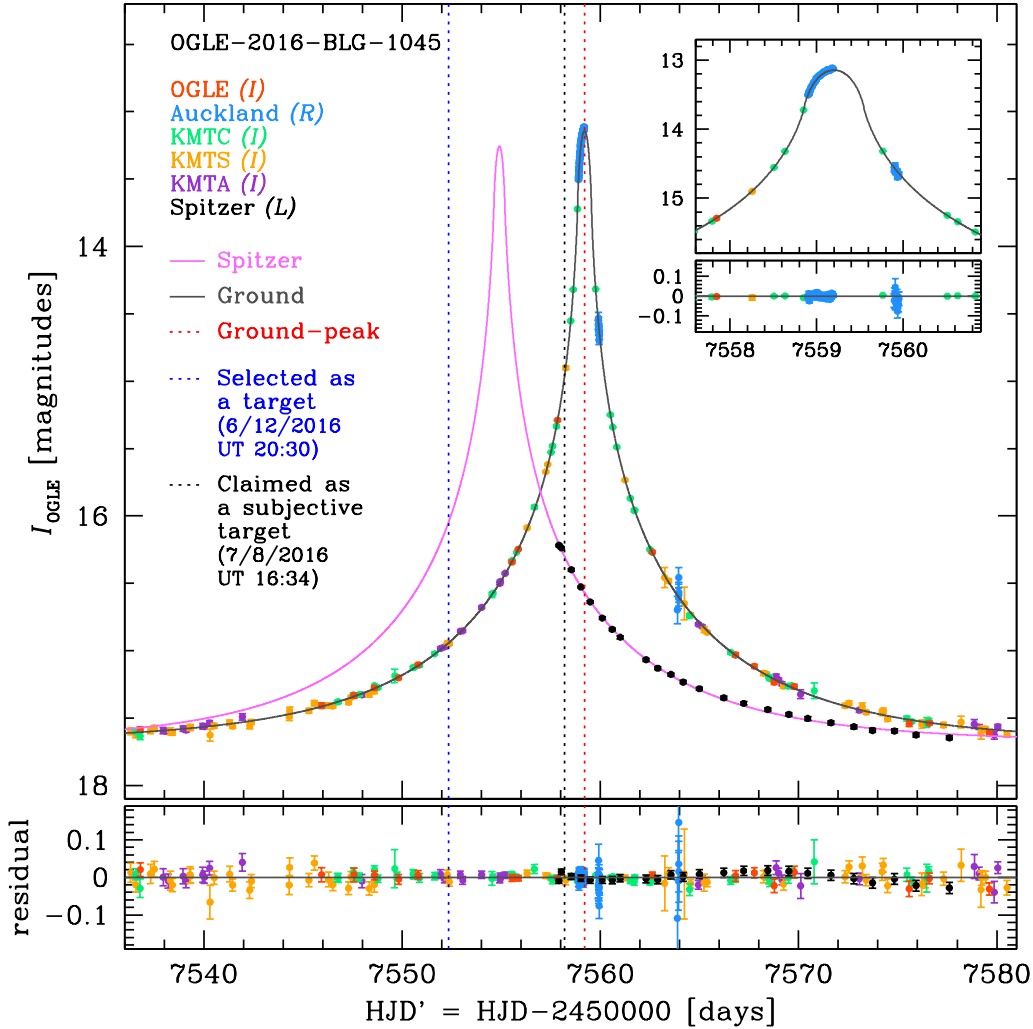}
\caption{
Light curves of the single-lens event OGLE-2016-BLG-1045 seen from the ground and space. Colored 
dots represent observed data taken from different telescopes located on the ground and in space 
(i.e., {\it Spitzer}). The dark gray and pink solid lines represent model light curves of the 
ground and {\it Spitzer}, respectively. The red dotted line indicates the peak time ($t_0$, see 
Table \ref{table:two}) of the ground-based light curve. The upper panel shows the observed light 
curves with their best-fit models. The lower panel shows residuals between the observations and 
the best-fit model. The inner panel shows the zoom-in of the peak part of ground-based light curve, 
which has a smooth feature due to the finite source effect. The dotted blue line indicates the time 
that this event was selected as a {\it Spitzer} target. The dotted black line indicates the time 
that the event was claimed as a {\it subjective} target.
\label{fig:one}}
\end{figure*}

\section{Test of the Cheap-SPRX Idea}

\subsection{Testbed: OGLE-2016-BLG-1045 {\it Spitzer} event}

\subsubsection{Ground Observations}

 The microlensing event OGLE-2016-BLG-1045 occurred on a source that lies at $(\alpha,\delta)_{\rm J2000}=
(17^{h}36^{m}51^{s}\\.19,-34^{\circ}32^{'}39^{''}.7)$, which corresponds to the Galactic coordinates $(l,b)
=(354.^{\circ}255,-1.^{\circ}386)$. The Optical Gravitational Lensing Experiment \citep[OGLE-IV:][]{udalski15a} found this 
event and then the Early Warning System \citep{udalski94,udalski03} of the OGLE-IV survey announced the event 
on $2016$ June $9$. The observations were made with the $1.3$ m Warsaw telescope in the $I-$band channel 
of a $1.4$ square-degree camera located at the Las Campanas Observatory in Chile.

 The event was highly magnified, implying that a planetary companion to the lens could probably 
be detected if it exists. Hence, a follow-up observation team called the Microlensing Follow-Up 
Network \citep[$\mu$FUN:][]{gould06} observed this event to capture any anomalies that might be 
produced by a planet. Auckland observatory, a $\mu$FUN member located in New Zealand, made the 
observations with a $0.4$ m telescope using a number $12$ Wratten filter (which is similar to $R-$band). 
The Auckland observations successfully covered the peak of the event. This peak coverage did not 
reveal an anomaly in the light curve due to a planetary lens system. However, the good coverage 
of the peak provided a chance to detect the finite source effect, which enters the determination of
the angular Einstein ring radius, i.e., $\rho_{\ast}=\theta_{\ast}/\theta_{\rm E}$.
The finite source effect can provide a mass-distance relation, $M/D_{\rm rel}=(c^2/4G)~\theta_E^2$, 
where the $D_{\rm rel}\equiv(D_{\rm L}^{-1}-D_{\rm S}^{-1})^{-1}$ is the relative distance between distances 
to the lens $(D_{\rm L})$ and the source $(D_{\rm S})$, $M$ is the lens mass, $c$ is the speed of light, 
and $G$ is the Newton's constant.

 There exist other $\mu$FUN observations in $H-$band taken at the Cerro Tololo International Observatory 
in Chile with the $1.3$ m SMARTS telescope (CTIO). These CTIO data were not included in the final models 
because of the similar coverage to the KMTNet data, but were used for the color-magnitude diagram 
(CMD) analysis of the event (see Appendix).

 The Korea Microlensing Telescope Network \citep[KMTNet:][]{kim16} also observed this event. Three identical 
$1.6$ m telescopes located in the Cerro Tololo International Observatory in Chile (KMTC), the South African 
Astronomical Observatory in South Africa (KMTS), and the Siding Spring Observatory in Australia (KMTA) 
observed this event with the $I-$band channel of their $4$ ${\rm deg}^{2}$ cameras. The KMTNet observations 
provided overall coverage of the light curve.

 The observed data sets were reduced by each group using their own pipelines and difference-imaging 
analysis packages: [\citep[OGLE-IV (DIA):][]{alard98,wozniak00}, \citep[$\mu$FUN and KMTNet (pySIS):]
[]{albrow09}.] 

\subsubsection{Space Observations}

\begin{figure*}[htb!]
\epsscale{0.90}
\plotone{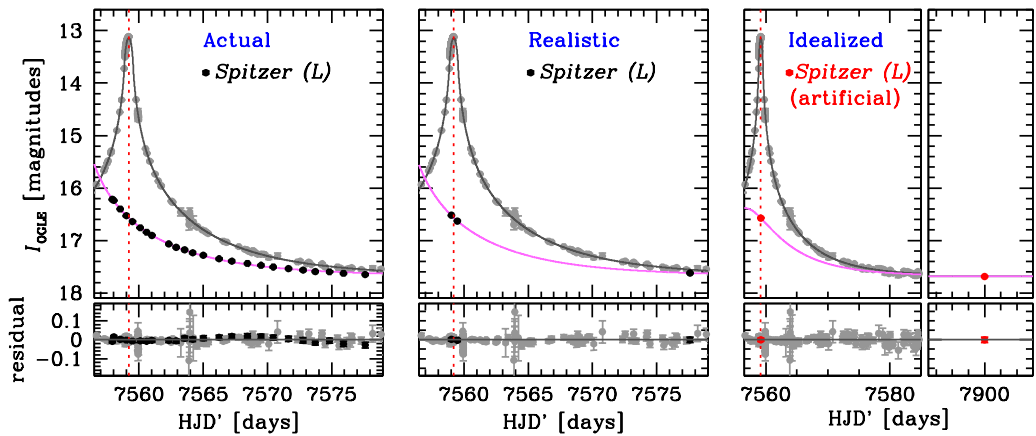}
\caption{
Light curves showing each test case. The left panel shows the ``Actual'' case using all actually 
observed $24$ {\it Spitzer} data points. The middle panel shows the ``Realistic'' case using $3$ 
selected {\it Spitzer} data points considering realistic space-based observations of the 
cheap-SPRX idea. The right panel shows the ``Idealized'' case using $2$ artificial data points 
considering the ideal situation of the cheap-SPRX idea. The gray, black, and red dots indicate 
ground-based observations, {\it Spitzer} observations, and the artificial data, respectively. 
Black and magenta lines represent the best-fit model light curves of $(-,+)$ solutions of each case.
\label{fig:two}}
\end{figure*}

 This event was {\it secretly} chosen as a target of the $2016$ {\it Spitzer Microlensing Campaign} 
on $2016$ June $16$ (UT $20$:$30$) based on the possibility that the event could be highly magnified. 
The event was later claimed as a ``subjective'' target on $2016$ June $18$ (UT $16$:$34$) once 
the event was observed to be moderate to high magnification (see \citealt{yee15b} for more details 
on different types of event selection). The observations began on $2016$ June $18$ (UT $9$:$56$) and 
ended on July $8$ (UT $2$:$43$). The {\it Spitzer Space Telescope} took $24$ total data points over 
$20$ days with the $3.6$ $\mu$m channel ($L-$band) of the IRAC camera. The {\it Spitzer} data were 
reduced with point response function photometry \citep{calchi15b}.

\subsubsection{Lightcurves}

 In Figure \ref{fig:one}, we present light curves of the event observed from ground and space. 
We also present the best-fit model lightcurves and their residuals, which is the $(-,+)$ 
case presented in Table \ref{table:two}. The ground-based light curve shows a symmetric 
Paczy\'nski curve \citep{paczynski86} with a smooth peak feature, which implies that the 
event was produced by a single lens affected by the finite source effect. The {\it Spitzer} 
observations only partially covered the light curve. However, \citet{han17}, \citet{shin17}, and 
\citet{wang17} already showed that it is possible to accurately measure the SPRX even though the 
space-based observations are fragmentary. Thus, for this event, using the {\it Spitzer} observations 
and the finite source effect, it is possible to measure the microlens parallax and the angular 
Einstein ring radius, which yield the properties of the isolated lens. We note that there 
exists a systematic trend in the {\it Spitzer} observations. The origin of this trend is unknown. 
However, several publications that used the {\it Spitzer} data with a similar trend 
\citep[e.g.,][]{poleski16,shin17,shvartzvald17,zhu17} concluded that the trend is not likely 
to affect determinations of their models. In this case, the trend is milder than those in 
the previous publications.

 The {\it Spitzer} observations were not taken with the idea of ``cheap-SPRX'' in mind. In fact, 
because the peak magnification was relatively unconstrained when the observations were scheduled, 
many similar events were observed on the chance that one of them would be high-magnification 
(so, these observations cannot be considered ``cheap''). Nevertheless, the resulting observations 
contain what would be obtained for a ``cheap-SPRX'' campaign, i.e., the {\it Spitzer} observations 
exist near the peak of the ground-based light curve and also exist near the baseline. Hence, this 
event can serve as an excellent testbed to perform a practical test of the cheap-SPRX idea.

\subsection{Test Method}

\subsubsection{Three Cases to Test the Cheap-SPRX Measurement}

 We test the accuracy of the cheap-SPRX method by considering three different {\it Spitzer} 
datasets, which we refer to as the ``Actual'', ``Realistic'', and ``Idealized'' cases. These datasets differ 
in the amount of information they contain (most to least). We first consider the two extremes, 
which are the ``Actual'' case defined by the current experiment and the ``Idealized'' GY12 case. 
For the ``Actual'' case, we use all observed {\it Spitzer} data ($24$ points). From this case, 
we can obtain the {\it actual} SPRX measurement that can be used as a reference to compare with 
the measurements derived from the other cases. For the ``Idealized'' case, considering the 
{\it ideal} situation proposed by GY12, this represents the minimum amount of data necessary 
for the cheap-SPRX idea to work. For this case, we generate two artificial data points using 
the {\it Spitzer} data and the best-fit model light curve. One is located at the exact 
ground-based peak (${\rm HJD'}=7559.201$) and the other is located at the baseline 
(${\rm HJD}'=7900.000$). For the ``Realistic'' case, we choose two actual data points near 
the ground-based peak (${\rm HJD}'=7559.172$ and $7559.482$) because it is almost impossible 
to take an image at the exact peak time in realistic situations. In addition, we use the last 
point (${\rm HJD}'=7577.613$) observed by {\it Spitzer}, which is located near the baseline. 
Based on these selected {\it Spitzer} data, we can obtain a measurement of the cheap-SPRX 
under {\it realistic} conditions. In Figure \ref{fig:two}, we present light curves of 
the cases that clearly show the space-based observations used for the test.

\subsubsection{Modeling of Lightcurves}

\begin{figure*}[htb!]
\epsscale{1.00}
\plotone{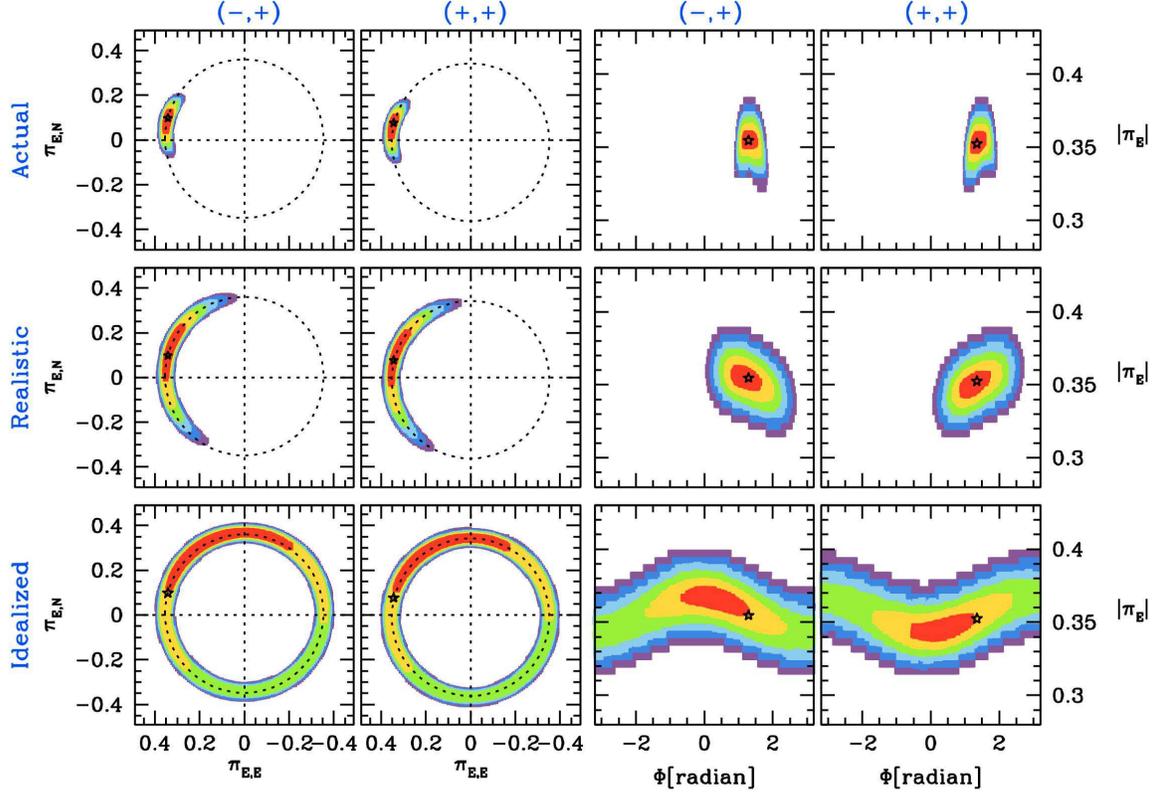}
\caption{
The SPRX distributions of each case with degenerate solutions. The top, middle, and bottom panels 
show the SPRX distributions of the actual, realistic, and idealized cases, respectively. The left 
six panels present the ($\pi_{{\rm E},E}$, $\pi_{{\rm E},N}$) distributions according to the 
conventional parameterization. The right six panels present the ($\pi_{\rm E}$, $\Phi$) 
distributions that are the MCMC parameters used to describe the SPRX. The red, yellow, green, 
light blue, blue, and purple colors represent $\Delta\chi^2=1^2$, $2^2$, $3^2$, $4^2$, $5^2$, 
and $6^2$, respectively. The star symbols indicate the best-fitted SPRX value of the actual case.
\label{fig:three}}
\end{figure*}

 Based on the three cases, we conduct modeling to measure the SPRX value of each case. For the modeling, 
we use six parameters: ($t_0$, $u_0$, $t_E$, $\rho_{\ast}$, $\pi_{\rm E}$, and 
$\Phi$). Among them, three basic parameters ($t_0$, $u_0$, and $t_E$) describe the light curve produced 
by a single-lens and a point-source. These basic parameters are closely related to each other: $t_0$ 
is the time at the peak of the light curve; $u_0$ is the impact parameter, i.e., the separation 
between the center of the Einstein ring and the position of the source at time $t_0$; $t_{\rm E}$ 
is the crossing-time of the Einstein ring. Another parameter $\rho_{\ast}$ is the angular source 
radius ($\theta_{\ast}$) normalized by the angular Einstein ring radius ($\theta_{\rm E}$), $\rho_{\ast} 
\equiv \theta_{\ast}/\theta_{\rm E}$, which describes the finite source effect. The last two parameters 
($\pi_{\rm E}$ and $\Phi$) describe the SPRX, which differs from the conventional way of describing 
the microlens parallax vector $\pivec$ (normally consisting of North ($\pi_{{\rm E},N}$) and East 
($\pi_{{\rm E},E}$) components). In our parameterization (see also \citealt{bennett08}),
\begin{equation}
\pivec_{\rm E} = (\pi_{{\rm E},N},\pi_{{\rm E},E}) \rightarrow (\pi_{\rm E}\cos\Phi,\pi_{\rm E}\sin\Phi).
\label{eqn:piereform}
\end{equation}
The $\Phi$ angle is allowed to vary over the full possible range $[-\pi, +\pi]$\footnote[2]{The parameter 
$\Phi$ is treated as a cyclic variable. That is, whenever it crosses the ``boundaries'' at $\pm \pi$, its 
formal value is changed by $\mp 2\pi$, so that there are no rejected links due to these ``boundaries''.}.
In addition, there are flux parameters ($F_{S}$ and $F_{B}$) for each data set that describe the fluxes of 
the source and blend, respectively, which are fit linearly for each model. We note that the model flux 
for each dataset, $i$, is derived from $F_{{\rm obs},i}(t)={\rm A}(t)F_{S,i}+F_{B,i}$, where the ${\rm A}(t)$ 
is the model magnification as a function of time.
Using these parameters, we search for the best-fit model with the minimum $\chi^2$ between the observed and 
modeled light curves using a Markov Chain Monte Carlo (MCMC) $\chi^2$ minimization 
\citep[the details of our MCMC sampling method are described in][]{dunkley05}. 
To find the global minimum of the model parameters, especially the SPRX parameter ($\pi_{\rm E}$), 
we initially conducted a grid search over $\pi_{\rm E}$ and $\Phi$ using the $200$x$200$ grid points. 
The grid search results are same as those of the MCMC simulations.

\begin{deluxetable}{llc}
\tablecaption{Limb-darkening coefficients and error re-scaling factors\label{table:one}}
\tablewidth{0pt}
\tablehead{
\multicolumn{1}{c}{Observations} &
\multicolumn{1}{c}{$\Gamma_{\lambda}$} & 
\multicolumn{1}{c}{$k$}
}
\startdata
OGLE ({\it I})                 & 0.5103        & 0.913 \\
Auckland ({\it R})$^{\dagger}$ & 0.6583        & 2.370 \\
KMTC ({\it I})                 & 0.5103        & 1.116 \\
KMTS ({\it I})                 & 0.5103        & 1.501 \\
KMTA ({\it I})                 & 0.5103        & $~$1.446 
\enddata
\tablecomments{
$^{\dagger}$We use a modified LD coefficient for Auckland observations, 
$\Gamma_R = (\Gamma_{R}+\Gamma_{V})/2= (0.61118 + 0.7048)/2 = 0.6583$ 
because the Auckland observatory used a $12$ Wratten filter having 
a flat transmission between $540-700$ nm. Thus, the filter is similar to 
the mean value of $R-$ and $V-$bands. 
Note that we did not use a $\Gamma_{L}$ because it plays no role for 
the {\it Spitzer} observations.
}
\end{deluxetable}

\begin{deluxetable*}{lrrrrrr}
\tablecaption{The best-fit model with degenerate solutions of each case\label{table:two}}
\tablewidth{0pt}
\tablehead{
\multicolumn{1}{c}{Case} &
\multicolumn{2}{c}{Actual} &
\multicolumn{2}{c}{Realistic} &
\multicolumn{2}{c}{Idealized} \\
\multicolumn{1}{c}{parameter} &
\multicolumn{1}{c}{$(-,+)$} &
\multicolumn{1}{c}{$(+,+)$} &
\multicolumn{1}{c}{$(-,+)$} &
\multicolumn{1}{c}{$(+,+)$} &
\multicolumn{1}{c}{$(-,+)$} &
\multicolumn{1}{c}{$(+,+)$} 
}
\startdata
$\chi^2_{\rm total} / N_{\rm data}$   & 1368.70 / 1372               & 1368.99 / 1372               &   1344.89 / 1351             &   1345.04 / 1351             &    1343.83 / 1350            &   1343.95 / 1350             \\
$\chi^2_{\rm Ground} / N_{\rm data}$  & 1345.01 / 1348               & 1345.09 / 1348               &   1344.77 / 1348             &   1344.77 / 1348             &    1343.83 / 1348            &   1343.95 / 1348             \\
$\chi^2_{\it Spitzer} / N_{\rm data}$ &   23.69 / 24                 &   23.90 / 24                 &      0.12 / 3                &      0.27 / 3                &       0.00 / 2               &      0.00 / 2                \\
$\chi^2_{\rm penalty}$                &    0.017                     &    0.075                     &      0.000                   &      0.010                   &       0.003                  &      0.014                   \\
$(I-L)$ [3.80]                        &    3.797                     &    3.794                     &      3.800                   &      3.802                   &       3.799                  &      3.803                   \\
$t_0$ (HJD')                          & 7559.201$\pm$0.001           & 7559.201$\pm$0.001           & 7559.201$\pm$0.001           & 7559.201$\pm$0.001           & 7559.202$\pm$0.001           & 7559.202$\pm$0.001           \\
$u_0$ ($10^{-2}$)                     &   -1.308$_{-0.042}^{+0.033}$ &    1.314$_{-0.044}^{+0.036}$ &   -1.318$_{-0.037}^{+0.041}$ &    1.318$_{-0.044}^{+0.033}$ &   -1.312$_{-0.044}^{+0.033}$ &    1.309$_{-0.033}^{+0.044}$ \\
$t_{\rm E}$ (days)                    &   11.981$_{-0.098}^{+0.064}$ &   11.963$_{-0.084}^{+0.088}$ &   11.950$_{-0.083}^{+0.084}$ &   11.947$_{-0.083}^{+0.088}$ &   11.956$_{-0.094}^{+0.073}$ &   11.952$_{-0.088}^{+0.076}$ \\
$\rho_{\ast}$ ($10^{-2}$)             &    3.186$_{-0.026}^{+0.033}$ &    3.190$_{-0.030}^{+0.030}$ &    3.195$_{-0.030}^{+0.027}$ &    3.195$_{-0.033}^{+0.028}$ &    3.194$_{-0.029}^{+0.030}$ &    3.193$_{-0.028}^{+0.031}$ \\
$\pi_{\rm E}$                         &    0.355$_{-0.006}^{+0.004}$ &    0.352$_{-0.005}^{+0.006}$ &    0.355$_{-0.008}^{+0.005}$ &    0.350$_{-0.006}^{+0.008}$ &    0.365$_{-0.015}^{+0.004}$ &    0.346$_{-0.004}^{+0.014}$ \\
$\Phi$ (radian)                       &    1.291$_{-0.062}^{+0.165}$ &    1.353$_{-0.066}^{+0.167}$ &    1.210$_{-0.284}^{+0.381}$ &    1.178$_{-0.177}^{+0.458}$ &    0.341$_{-1.185}^{+0.955}$ &    0.407$_{-1.188}^{+1.141}$ \\
$\pi_{{\rm E},{\it E}}$               &    0.341$_{-0.012}^{+0.012}$ &    0.344$_{-0.011}^{+0.013}$ &    0.332                     &    0.323                     &    0.122                     &    0.137                     \\
$\pi_{{\rm E},{\it N}}$               &    0.098$_{-0.059}^{+0.027}$ &    0.076$_{-0.058}^{+0.028}$ &    0.125                     &    0.134                     &    0.344                     &    0.317                     \\
$F_{\rm S, OGLE}$                     &    1.370$_{-0.010}^{+0.014}$ &    1.373$_{-0.013}^{+0.012}$ &    1.375$_{-0.012}^{+0.012}$ &    1.375$_{-0.013}^{+0.011}$ &    1.374$_{-0.011}^{+0.014}$ &    1.374$_{-0.012}^{+0.013}$ \\
$F_{\rm B, OGLE}$                     &   -0.032$_{-0.014}^{+0.010}$ &   -0.034$_{-0.012}^{+0.012}$ &   -0.036$_{-0.012}^{+0.011}$ &   -0.037$_{-0.012}^{+0.012}$ &   -0.035$_{-0.014}^{+0.010}$ &   -0.036$_{-0.013}^{+0.011}$ \\
$F_{\rm S, {\it Spitzer}}$            &   45.257$_{-0.932}^{+0.919}$ &   45.212$_{-0.870}^{+1.004}$ &   45.518$_{-1.061}^{+0.953}$ &   45.622$_{-1.199}^{+0.902}$ &   45.439$_{-0.909}^{+1.124}$ &   45.618$_{-1.127}^{+0.973}$ \\
$F_{\rm B, {\it Spitzer}}$            &   -6.528$_{-0.993}^{+0.941}$ &   -6.430$_{-1.178}^{+0.819}$ &   -8.032$_{-1.094}^{+0.963}$ &   -8.179$_{-1.046}^{+1.174}$ &   -6.674$_{-1.190}^{+0.844}$ &   -6.854$_{-1.039}^{+1.062}$ \\
\enddata
\tablecomments{
${\rm HJD' = HJD-2450000.0}$. 
The $N_{\rm data}$ after each $\chi^2$ value indicates the number of data points that are used for the modeling.
We note that the $\pi_{{\rm E},E}$ and $\pi_{{\rm E},E}$ are not modeling parameters. 
These are calculated from the modeling parameters, $\pi_{\rm E}$ and $\Phi$ (see Equation~(\ref{eqn:piereform})).
We do not present the errors of $\pi_{{\rm E},N}$ and $\pi_{{\rm E},E}$ for Realistic and Idealized cases because 
these errors are meaningless: only the error in $\pi_{\rm E}$ has meaning.
}
\end{deluxetable*}

 During the modeling process, we consider the limb-darkening (LD) of the source star. We adopt LD 
coefficients for observed passbands from \citet{claret00} based on the spectral source type determined 
by the CMD analysis (described in the Appendix). In addition, we re-scale the errors of observations 
to enforce $\chi^2/{\rm dof}\simeq 1$ using the equation $e_{\rm new}= k(e_{\rm old})$ where $k$, 
$e_{\rm new}$, and, $e_{\rm old}$ are the error re-scaling factor, re-scaled errors, and original 
errors, respectively. The error re-scaling process has been done based on the best-fit model, i.e., 
the $(-,+)$ case. We note that, in the case of the OGLE-IV data, the observational errors are 
calibrated using a correction procedure that is described in \citet{skowron16}, before applying 
the error re-scaling process based on the best-fit model. In Table \ref{table:one}, we present these 
LD coefficients and error re-scaling factors for modeling.
 
 We also incorporate the color-constraint, $(I-L)=3.800\pm0.020$, which provides an independent 
constraint on the model. The constraint is determined using $I-$band ground observations (OGLE-IV) 
and $L-$band space observations ({\it Spitzer}) based on the CMD analysis. To incorporate the 
($I-L$) color-constraint, we introduce $\chi^2_{\rm penalty}$ described in Section 3.2 of 
\citet{shin17}. The $\chi^2_{\rm penalty}$ increases the $\chi^2$ when the fitted ($I-L$) color 
of the model is different from the constraint. In particular, the $\chi^2_{\rm penalty}$ increases 
strongly when the difference between the fitted color and the constraint is larger than $2\sigma$. 

 In Table \ref{table:two}, we present the best-fit parameters for each case (Actual, Realistic, Idealized). 
For each case, we find that there exist two degenerate solutions due to the ``four-fold degeneracy'' 
\citep{refsdal66,gould94}. In principle, the four-fold degeneracy has four solutions, $(+,+)$, $(+,-)$, 
$(-,+)$, and $(-,-)$ (denoted according to the convention described in \citealt{zhu15}), which are caused by 
different pairs of source trajectories (seen from ground and space) going through a similar 
lensing magnification pattern. This degeneracy can be divided into two categories by its origin 
(GY12 and references therein). The first (denoted by the first $\pm$ sign in this paper) is related to the relative 
positions of the Earth and satellite, whether they lie on the same or opposite sides of the lens. The other (denoted 
by the second $\pm$ sign in this paper) is related to the different possible source trajectories as seen from Earth, 
i.e., whether they pass on the left or right sides of the lens. The former degeneracy can affect the magnitude ($\pi_{\rm E}$) 
of the $\pivec_{\rm E}$, while the latter degeneracy can only affect the direction of the $\pivec_{\rm E}$, 
which is less interesting in this test of the cheap-SPRX idea. The four-fold degeneracy can sometimes be resolved 
\citep[e.g.,][]{chung17, han16, han17, shin17, udalski15b, yee15a}. For this event, we find that there exist 
only two solutions, $(-,+)$ and $(+,+)$, based on the grid search process. The other two solutions, 
$(-,-)$ and $(+,-)$, are merged with the $(-,+)$ and $(+,+)$ solutions, respectively. 
The reason that the four solutions are merged into only two solutions for this event is that $u_{0,Spitzer}\sim 0$. 
For model parameters of each solution, uncertainties are determined based on the $68\%$ confidence 
intervals of the MCMC chains.

\begin{figure*}[htb!]
\epsscale{0.90}
\plotone{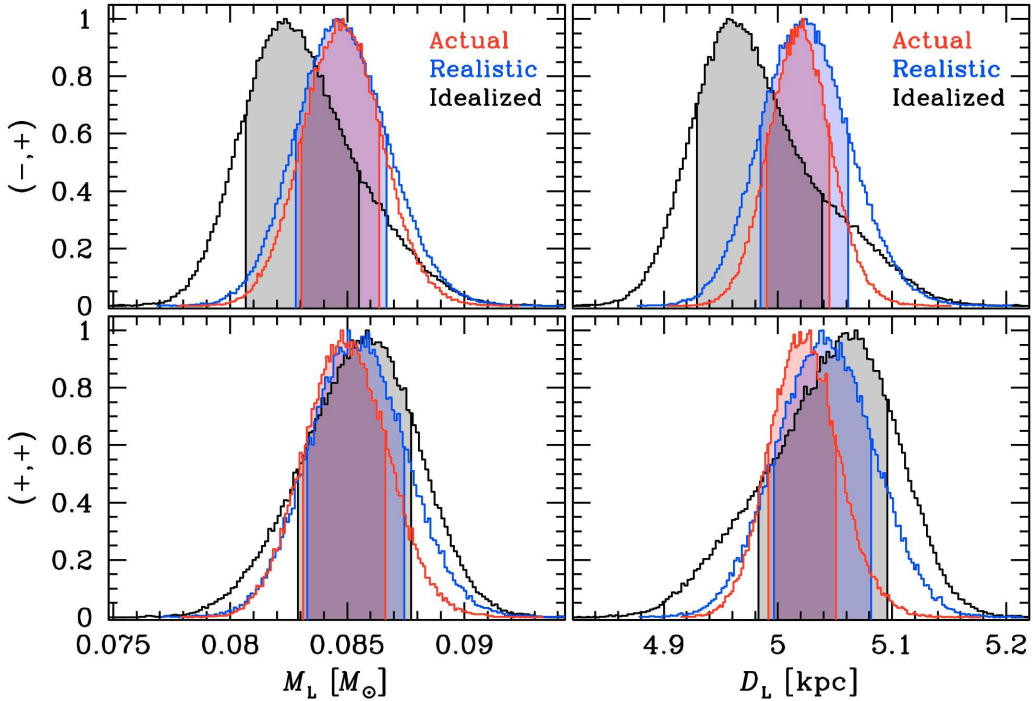}
\caption{
Distributions of lens properties determined from the MCMC chains.
The left-side upper and lower panels show the lens mass distributions of $(-,+)$ and $(+,+)$ solutions, 
respectively. The right-side upper and lower panels show the distributions of the distance to the lens 
of $(-,+)$ and $(+,+)$ solutions, respectively. The red, blue, and black colors indicate the actual, 
realistic, and idealized case, respectively. The colored shade shows the $1\,\sigma$ uncertainty ($68\%$ 
area of the distributions) of each case. Each distribution is normalized so the peak of the histogram is 
set to unity.
\label{fig:four}}
\end{figure*}

\subsection{Test Results}

\subsubsection{Validation of the Accuracy of the Cheap-SPRX Measurement}

 In Figure \ref{fig:three}, we present the SPRX distributions of each case. The distributions are 
constructed from the MCMC chains. These distributions clearly show the consistency of the SPRX 
measurements. We present two types of distributions. One type of distribution is presented according 
to the conventional parameters, $(\pi_{{\rm E},E}, \pi_{{\rm E},N})$, which are calculated from the MCMC parameters 
as $\pi_{{\rm E},E}=\pi_{\rm E}\sin{\Phi}$ and $\pi_{{\rm E},N}=\pi_{\rm E}\cos{\Phi}$. The other 
is the ($\pi_{\rm E}$, $\Phi$) distribution, which can be used to directly check the accuracy of the 
magnitude of the SPRX measurement. 

 From the modeling of the actual case, we obtain the SPRX measurements for the $(-,+)$ and $(+,+)$ 
cases: $\pi_{\rm E}=0.355^{+0.004}_{-0.006}$ and $0.352^{+0.006}_{-0.005}$, respectively. 
We find that the magnitudes of the SPRX values between the $(-,+)$ and $(+,+)$ solutions of the actual 
case are consistent to well within $1\sigma$. Based on the actual SPRX measurements, we can compare 
the other test cases of the cheap-SPRX idea to check the accuracy of the cheap-SPRX measurements. 
For the realistic case, we find that the SPRX measurements of both degenerate solutions, 
$0.355^{+0.005}_{-0.008}$ and $0.350^{+0.008}_{-0.006}$, are consistent with those of the actual case 
to within $1\sigma$. 
For the idealized case, the measurements, $0.365^{+0.004}_{-0.015}$ and $0.346^{+0.014}_{-0.004}$, are
consistent to within $\lesssim 1 \sigma$ using the idealized-case errors.

 Based on the SPRX measurements, we can determine the properties of this isolated lens by combining 
it with the angular Einstein ring radius ($\theta_{\rm E}=\theta_{\ast}/\rho_{\ast}$), where 
$\theta_{\ast}$ is the angular source radius determined from the CMD analysis (described in the 
Appendix) and $\rho_{\ast}$ is determined from the finite source effect. We determine the angular 
Einstein ring radius as
\begin{equation}
\theta_{\rm E} = 0.244 \pm 0.015 ~{\rm mas}.
\end{equation}

 In Figure \ref{fig:four}, we present distributions of physical properties of the lens for each case. 
the lens mass ($M_{\rm L}$) and the lens distance ($D_{\rm L}$) are determined from MCMC parameters as
\begin{equation}
M_{\rm L} = {(\theta_{\ast}/\kappa) \over {\rho_{\ast}\pi_{\rm E}}},~~\kappa = 8.144 ~{\rm mas}\, M_{\odot}^{-1},
\end{equation}
\begin{equation}
D_{\rm L} ={ {\rm au} \over {(\pi_{\rm E}/\rho_{\ast})\,\theta_{\ast} + \pi_{S}} }, ~~\pi_{S} ={ {\rm au} \over D_{S} },
\end{equation}
where $D_{S}$ is the distance to the source estimated from \citet{nataf13}. 
For this event, the estimated $D_{S}$ is $\sim 8.87\,$ kpc. We find that both properties are consistent 
to within $1\sigma$ across all cases. In fact, the uncertainty in the properties is dominated by the 
uncertainty of the $\theta_{\ast}$ determination. Quantitatively, the uncertainty of the SPRX measurement 
is $< 3\%$ compared to the $\ge 6\%$ uncertainty in $\theta_{\ast}$. Thus, we find that the accuracy of 
the SPRX measurement based on the cheap-SPRX idea is sufficient to accurately determine the properties 
of the isolated object. The isolated lens of this event is a low-mass stellar object with $M_{\rm L}\sim 
0.08\pm0.01\, M_{\odot}$, which is located at $\sim 5.02\pm0.14\,$ kpc from us \footnote[3]{These values 
of physical properties are the simple mean values of each property, with the uncertainty determined 
through standard error propagation.}.

\begin{figure*}[htb!]
\epsscale{0.90}
\plotone{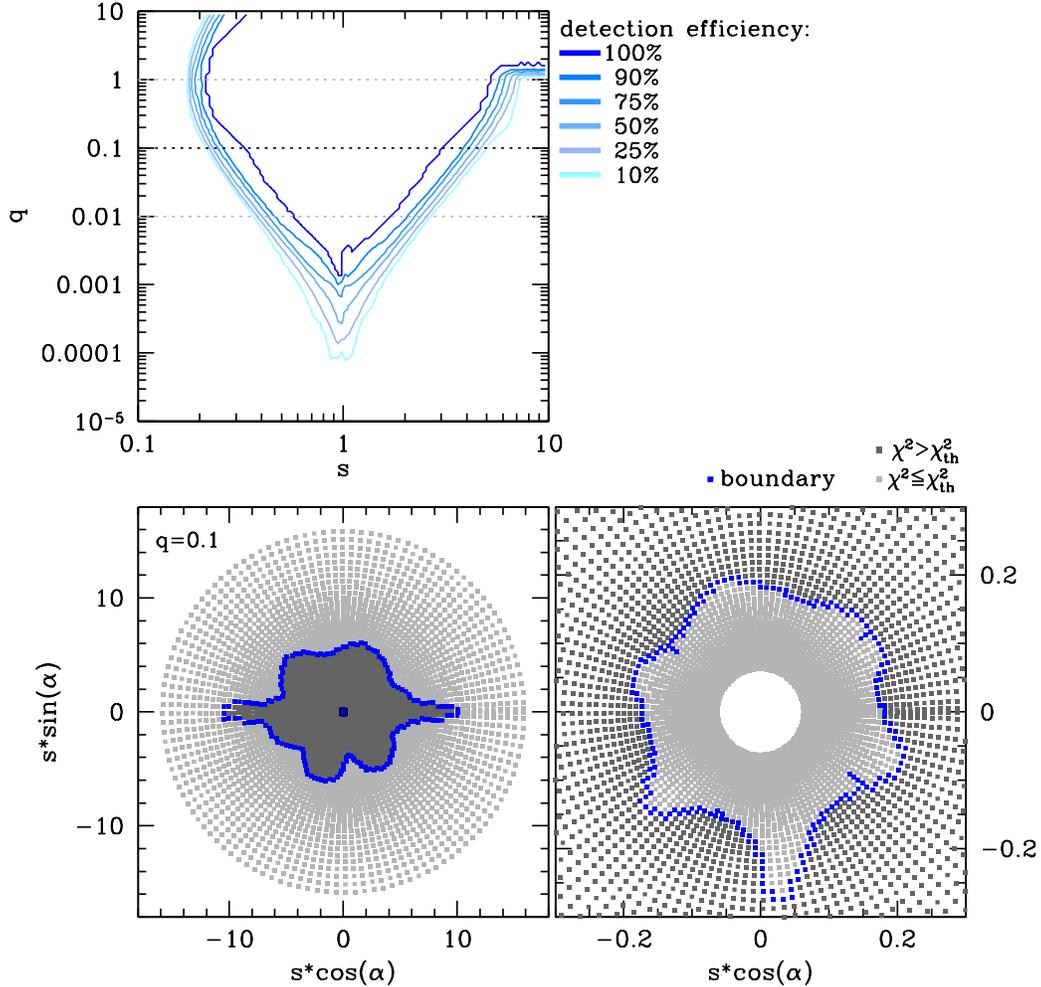}
\caption{
Example of diagrams at the mass ratio ($q=0.1$) and their boundaries. The upper panel shows 
the detection efficiency diagram built using $\chi^2_{\rm th}=15.0$. The lower panels show 
diagrams of the two regimes of binary lensing for the case of $q=0.1$. The left panel shows 
the wide ($s>1$) binary regime and the right panel shows the close ($s<1$) binary regime. 
The grey and dark grey dots represent two categories of binary-lensing cases whose boundary 
is given by $\chi^2_{\rm th}$. The grey dots indicate $\chi^2 \le 15.0$, while the dark grey 
dots indicate $\chi^2 > 15.0$. The blue dots indicate the boundary points between the two 
categories.
\label{fig:five}}
\end{figure*}

\begin{figure*}[htb!]
\epsscale{0.90}
\plotone{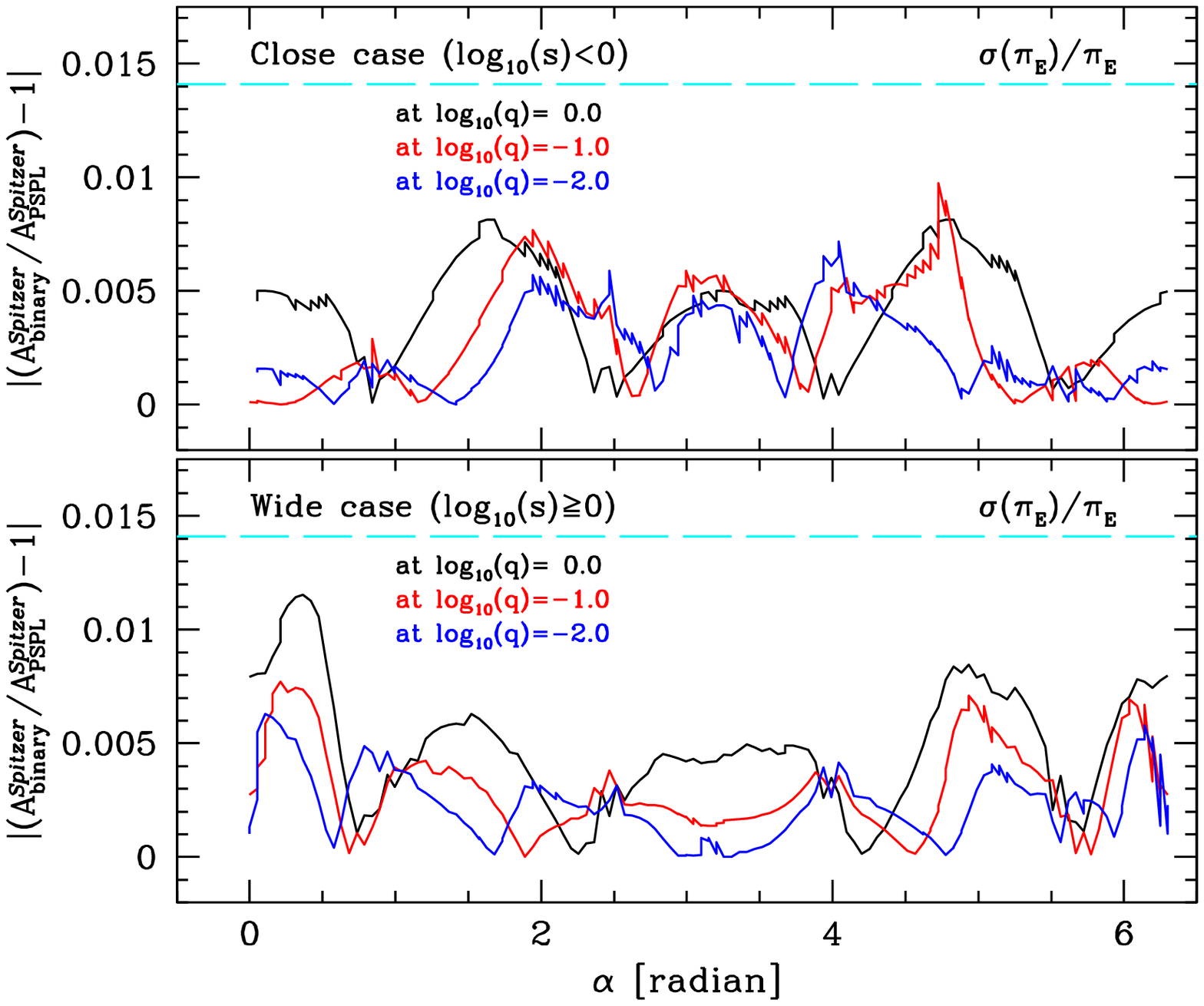}
\caption{
The criterion values of the test from the boundary cases.
The upper and lower panels show the criterion values for the close and wide binary cases, 
respectively. The black, red, and blue colors represent the magnification deviations of boundaries 
at $q=0.01,\,0.1,$ and $1.0$, respectively. The cyan dashed line represents the relative error of 
the measured SPRX value.
\label{fig:six}}
\end{figure*}

\subsubsection{Validation of Effects on the Cheap-SPRX Measurement by Binary-lensing Cases}

 The cheap-SPRX idea assumes that an observed lightcurve seen from space, e.g., 
the {\it Spitzer} observations, resembles a single-lensing lightcurve. However, 
if the lens is a binary and there are only two observations from the spacecraft, 
it will not be possible to determine from the space-based observations alone 
whether these are affected by the binary or whether the single-lens assumption 
is sufficient. Indeed, if the binary is not detected in the ground-based data, 
an anomaly in the space-based data due to a binary would go undetected. Then, 
the magnification computation to measure the cheap-SPRX may be inaccurately 
determined due to the effect of the binary-lensing perturbation on the lightcurve. 
As a result, a violation of the single-lensing assumption can in principle yield 
an incorrect measurement of the cheap-SPRX when a second mass exists.

However, high-magnification events (this is a basic assumption for applying 
the cheap-SPRX idea) are very sensitive to binary lenses. This implies that, 
for a high-magnification event, we can rule out a very broad class binary-lens 
configurations because these would produce clear anomalies on the ground-based 
lightcurve. We perform a quantitative test to check the effect on the cheap-SPRX 
measurement caused by binary-lensing. The test is performed using the following 
procedures.

 First, we separately conduct a binary-lens modeling with ground-based 
observations only. The best-fitting of this modeling yields a $\chi^2$ threshold 
to exclude binary-lensing cases, which have noticeable anomalies. The best-fit 
model has $\Delta\chi^2=(\chi^2_{\rm single}-\chi^2_{\rm binary})=13.9$.
Thus, we set the $\chi^2$ threshold $\chi^2_{\rm th}=15.0$. This is the criterion 
for dividing simulated binary-lensing cases into two categories: $\chi^2 > 
\chi^2_{\rm th}$ are the cases with anomalies that are detectable in the ground-based 
lightcurve, and $\chi^2 < \chi^2_{\rm th}$ are the cases having non-detectable anomalies.

 Second, we simulate binary-lensing cases with only ground-based observations 
using the Rhie method \citep{bennett96, rhie00}. In this procedure, 
the binary-lensing cases are simulated using a grid of the projected separation 
($s$), mass ratio ($q$), and angle ($\alpha$) of the source trajectory with 
respect to the binary-axis: $\log{s}=[-1.2,\, 1.2]$, $\log{q}=[-5.0,\, 1.0]$, 
and $\alpha=[0,\, 2\pi]$. Each range of the grid is divided into $120$ grid points (i.e., 
total $120^{3}$ binary-lensing cases are simulated). We adopt the other parameters, 
$t_0$, $u_0$, $t_{\rm E}$, and $\rho_{\ast}$, from the actual $(-,+)$ solution to 
produce an artificial dataset of the binary-lensing case. For each binary-lensing 
case with the artificial ground-based dataset, we calculate a $\chi^2$ value by 
fitting with a finite-source single-lensing model. 

 Third, we can build two types of diagrams (Figure \ref{fig:five}) using the simulated 
binary-lensing cases and the $\chi^2$ threshold: one is the diagram showing the detection 
efficiency of this event, and the other is the diagram showing two categories of 
the binary-lensing cases at a specified mass ratio. From this diagram, we can extract 
a ``boundary'' with $\Delta\chi^2=15$, which represents a kinds of extreme binary-lensing cases having 
non-detectable anomalies that may possibly affect the cheap-SPRX measurement. 
In Figure \ref{fig:five}, we present an example of such diagrams at the $q=0.1$ and their boundaries.

 Fourth, at these boundary cases, we can check the effect on the cheap-SPRX measurement caused 
by the hidden anomalies of the binary-lensing cases. To quantitatively check the effect, we set 
a criterion as
\begin{equation}
\left|\frac{A_{\rm binary}^{Spitzer}}{A_{\rm PSPL}^{Spitzer}} -1 \right|_{\rm peak,\oplus} < 
\frac{\sigma(\pi_{\rm E})}{\pi_{\rm E}}
\label{eqn:criterion}
\end{equation}
where the $A_{\rm binary}^{Spitzer}$ and $A_{\rm PSPL}^{Spitzer}$ are magnifications of the {\it Spitzer} lightcurve 
at the ground-peak time (HJD'$\sim7559.20$) computed using binary-lens and single-lens models, 
respectively. The $\pi_{\rm E}$ and $\sigma(\pi_{\rm E})$ are the cheap-SPRX measurement and its 
uncertainty adopted from the actual $(-,+)$ case. 
This criterion shows how much an undetected anomaly due to binary-lensing could affect the 
magnification of the {\it Spitzer} lightcurve. If the criterion in Equation~(\ref{eqn:criterion}) 
is met, the inaccuracy in the magnification is less significant than uncertainties from other 
sources. Using this criterion, we check three cases of boundaries at $q=0.01,\,0.1,$ and $1.0$.

 In Figure \ref{fig:six}, we present the quantitative results of this test. We find that, for all 
cases along the boundary, the deviations between magnifications of the {\it Spitzer} lightcurve at 
the ground-peak are much smaller than the relative error of the SPRX that is actually measured. 
This implies that the binaries that do not give to detectable signals in the ground-based data 
also do not significantly affect the SPRX measurement. 
Hence, in this case, even if there exists an undetected binary-lens anomaly, we can still obtain 
an accurate SPRX measurement using the cheap-SPRX idea.

\section{Conclusion and Discussion}

 Based on the event OGLE-2016-BLG-1045, we tested the cheap-SPRX idea to check the accuracy of 
the microlens parallax measurement by comparing it to the true measurement. In addition, based on 
the parallax measurement of each case, we checked whether the physical properties of this 
isolated lens are consistent or not. We found that the magnitudes of the actual SPRX measurement and the 
realistic, cheap-SPRX measurement are consistent to within $1\sigma$. We also found 
that the lens mass determined for all cases is consistent $\sim 0.08$ $M_{\odot}$, which is the 
upper-mass limit for brown dwarfs. In addition, the lens distances derived for all cases are also 
consistent to within $1\sigma$.
Moreover, we conducted a test to see how a binary lens that is not detectable in ground-based 
observations might affect the cheap-SPRX measurement. We found that this effect is not significant 
in this case.
Hence, we conclude that the cheap-SPRX measurement has sufficient accuracy 
to adopt this idea in real situations. Thus, using only two or three space-based observations, we 
can determine the physical properties of the lens for high-magnification events. This fact implies 
that by adopting the cheap-SPRX idea, we have a robust method of measuring microlens parallaxes 
(i.e., SPRX), which can reveal the nature of the lens with a cost-effective space-based campaign.

 A space-based microlensing campaign, perhaps added on to another mission, adopting this cost-effective 
idea can provide a measurement of the magnitude of the microlens parallax for most high-magnification 
events. This complete sample can be used to study isolated objects, especially low-mass objects, 
in the Galaxy and derive a mass function based on them. 

\mbox{}

\acknowledgments 
This research has made use of the KMTNet system operated by the Korea Astronomy and Space Science 
Institute (KASI) and the data were obtained at three host sites of CTIO in Chile, SAAO in South 
Africa, and SSO in Australia. 
This work is based in part on observations made with the Spitzer Space Telescope, which is 
operated by the Jet Propulsion Laboratory, California Institute of Technology under a contract 
with NASA.
OGLE project has received funding from the National Science Centre, Poland, grant MAESTRO 
2014/14/A/ST9/00121 to A. Udalski.
Work by I-G. Shin and A. Gould was supported by JPL grant 1500811.
A. Gould, Y. K. Jung, and W. Zhu acknowledges the support from NSF grant AST-1516842.
Work by YS and CBH was supported by an appointment to the NASA Postdoctoral Program at 
the Jet Propulsion Laboratory, administered by Universities Space Research Association 
through a contract with NASA. 
Work by C.H. was supported by the grant (2017R1A4A1015178) of National Research Foundation of Korea

\appendix
\section{The Color-Magnitude diagram (CMD) Analysis}

 From this CMD analysis, we can determine the angular source radius, the spectral type of 
the source star, and the model-independent color constraint. The CMD analysis is usually 
conducted by combining the $(V-I, I)$ CMD and the standard method \citep{yoo04}. However, 
for this event, the source is severely extincted with $A_I \sim 3.5$ in $I-$band. As a result, 
the standard method cannot be applied using the $(V-I,I)$ CMD. Hence, we construct a new 
$(I-H, I)$ CMD based on the OGLE-IV survey and the VISTA Variables and Via Lactea Survey 
\citep[VVV:][]{minniti10} using cross-matching of field stars, which are located within 
$60''$ from the source star.

\begin{figure}[htb!]
\epsscale{0.90}
\plotone{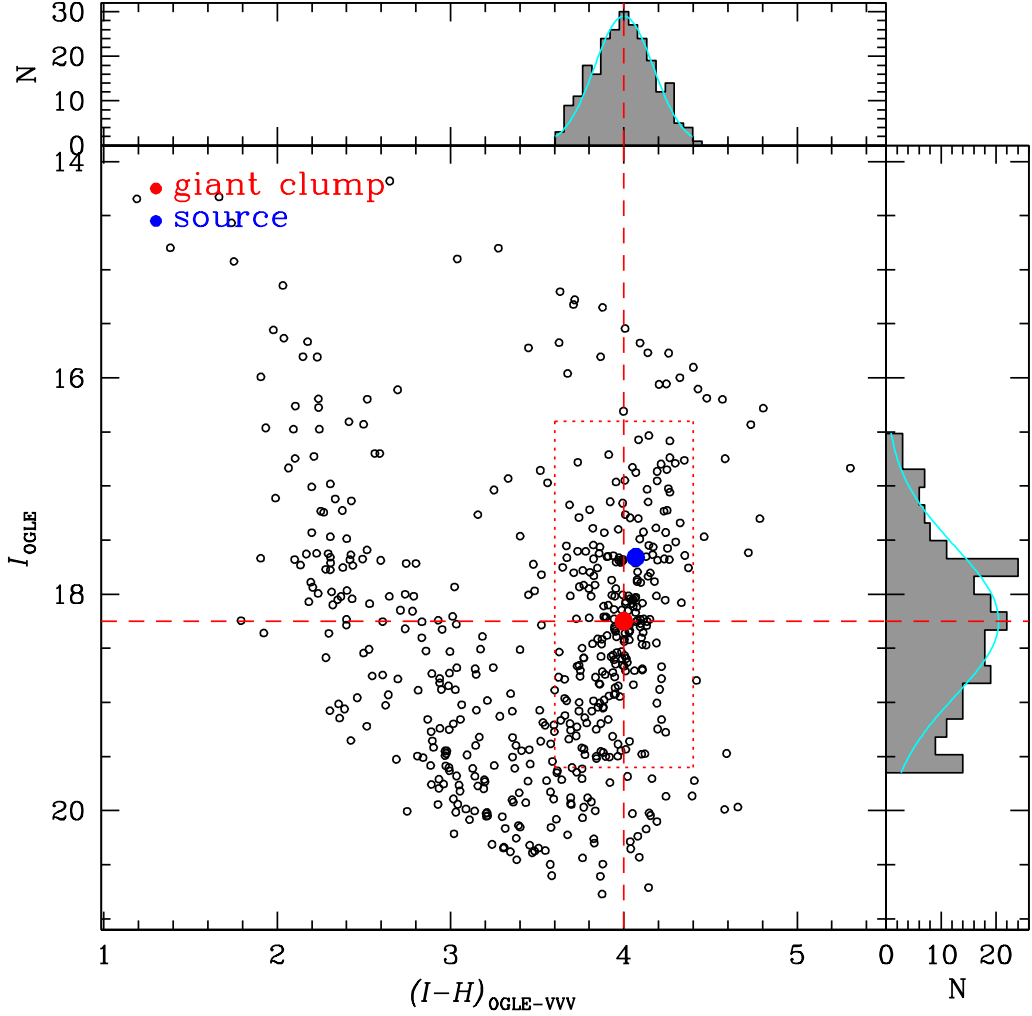}
\caption{
The $(I-H, I)$ CMD of the OGLE-2016-BLG-1045 event. The CMD is constructed by cross-matching OGLE-IV 
and VVV observations. The red and blue dots indicate the red giant clump centroid and the source,
respectively. We present color and magnitude distributions of field stars within a selected region, 
which is a box marked in red dotted lines, along with the abscissa and ordinate, respectively. The cyan 
line indicates the Gaussian fitting of the distributions. The location of the red giant clump centroid 
is determined from these distributions.
\label{fig:app1}}
\end{figure}

 In Figure \ref{fig:app1}, we present the $(I-H, I)$ CMD. We conduct the CMD 
analysis using the standard method. First, we determine the location of the red giant clump centroid 
on the CMD as $(I-H, I)_{\rm C} = (4.00\pm0.03, 18.25\pm0.05)$. Second, the location of the 
source on the CMD is determined based on source fluxes in $I$ band and $H$ band from the best-fit 
model additionally including CTIO $H-$band data. The magnitudes are found to be 
$I_{S,{\rm OGLE}}=17.658\pm0.004$ and $H_{S,{\rm CTIO}}=17.648\pm0.003$. The CTIO $H-$magnitude 
scale is converted to the VVV $H-$magnitude scale using the relation 
$(H_{\rm CTIO} - H_{\rm VVV})_{\rm S} = 4.059\pm0.011$, which comes from comparison stars. 
Thus, the location of the source on the CMD is determined to be 
$(I-H, I)_{\rm S} = (4.068\pm0.012,17.658\pm0.004)$.

 We adopt the de-reddened color \citep{bensby13} and intrinsic magnitude \citep{nataf13} of 
the giant clump as a reference. The adopted values are 
$(V-I, I)_{0,{\rm C}}=(1.06\pm0.01, 14.62\pm0.04)$. Based on this reference, we can obtain the 
de-reddened color and magnitude of the source under the assumption that the clump and source 
experience the same extinction. In addition, the $(I-H)$ color is converted to the $(V-I)$ 
color using the color-color relation in \citet{bessell88}. For the source of this event, the 
relation is $\Delta (I-H) = 1.00\times \Delta (V-I)$. Thus, the de-reddened color and magnitude 
of the source are $(V-I)_{0,{\rm S}} = (V-I)_{0,{\rm C}} - [(I-H)_{\rm C}-(I-H)_{\rm S}]$ 
and $I_{0,{\rm S}}=I_{0,{\rm C}} - [I_{\rm C} - I_{\rm S}]$, respectively. Lastly, we obtain 
the de-reddened color and magnitude of the source: $(V-I, I)_{0,{\rm S}} =(1.128\pm0.034, 14.028\pm0.064)$.

 From the color of the source, we determine the angular source radius using the color/surface-brightness 
relations in \citet{kervella04}. To employ the relation, we convert the $(V-I)_{0,{\rm S}}$ to 
$(V-I)_{0,{\rm S}}$ by using the \citet{bessell88} relation. The determined angular source radius is 
\begin{equation}
\theta_{\ast} = 7.80 \pm 0.47 ~~{\rm {\mu}as}.
\end{equation}

Moreover, based on the intrinsic source color, we estimate the source star to be an early K-type giant. 
We adopt LD coefficients from \citet{claret00} assuming typical properties of an early K-type giant: 
effective temperature $T_{\rm eff} \simeq 4750$ K, surface gravity $\log{g} \simeq 2.0$, 
microturbulent velocity $V_{t} \simeq 2.0$ ${\rm km\,s^{-1}}$, and metallicity $\log{[M/H]} \simeq 0.0$. 
The adopted LD coefficients are presented in Table \ref{table:one}.

Based on the information of the source, we determine the $(I-L)$ color constraint using the color-color 
regression method based on the $IHL$ color-color diagram. This process is described in \citet{calchi15b} 
and \citet{shin17}. The determined $(I-L)$ color constraint is 
\begin{equation}
(I-L)=3.800 \pm 0.020.
\end{equation}
We incorporate this model-independent constraint in the modeling process by introducing an additional 
$\chi^2_{\rm penalty}$, which increases as $\Delta (I-L)$ increases between the color calculated from 
the model and the constraint.



\begin{thebibliography}{999}

\bibitem[Alard \& Lupton (1998)]{alard98} 
Alard, C. \& Lupton, Robert H.\ 1998, \apj, 503, 325

\bibitem[Albrow et al.(2009)]{albrow09} 
Albrow, M. D., Horne, K., Bramich, D. M., et al.\ 2009, \mnras, 397, 2099

\bibitem[Alcock et al.(1995)]{alcock95} 
Alcock, C., Allsman, R.~A., Alves, D., et al.\ 1995, \apjl, 454, L125 

\bibitem[Barclay et al.(2017)]{barclay17} 
Barclay, T., Quintana, E.~V., Raymond, S.~N., et al.\ 2017, \apj, 841, 86 

\bibitem[Bennett et al.(2008)]{bennett08}
Bennett, D.P., Bond, I.A., Udalski, A., et al.\ 2008, \apj, 684, 663

\bibitem[Bennett \& Rhie(1996)]{bennett96} 
Bennett, D.~P., \& Rhie, S.~H.\ 1996, \apj, 472, 660 

\bibitem[Bensby et al.(2013)]{bensby13}
Bensby, T., Yee, J. C., Feltzing, S., et al.\ 2013, A\&A, 549, 147

\bibitem[Bessell \& Brett(1988)]{bessell88}
Bessell, M. S., \& Brett, J. M.\ 1988, \pasp, 100, 1134

\bibitem[Bihain et al.(2009)]{bihain09} 
Bihain, G., Rebolo, R., Zapatero Osorio, M.~R., et al.\ 2009, \aap, 506, 1169 

\bibitem[Bozza et al.(2016)]{bozza16} 
Bozza, V., Shvartzvald, Y., Udalski, A., et al.\ 2016, \apj, 820, 79 

\bibitem[Calchi Novati et al.(2015a)]{calchi15a} 
Calchi Novati, S., Gould, A., Udalski, A., et al.\ 2015a, \apj, 804, 20

\bibitem[Calchi Novati et al.(2015b)]{calchi15b} 
Calchi Novati, S., Gould, A., Yee, J. C., et al.\ 2015b, \apj, 814, 92

\bibitem[Chatterjee et al.(2008)]{chatterjee08} 
Chatterjee, S., Ford, E.~B., Matsumura, S., et al.\ 2008, \apj, 686, 580 

\bibitem[Chung et al.(2017)]{chung17} 
Chung, S.-J., Zhu, W., Udalski, A., et al.\ 2017, \apj, 838, 154 

\bibitem[Claret (2000)]{claret00}
Claret, A.\ 2000, A\&A, 363, 1081

\bibitem[Dong et al.(2007)]{dong07} 
Dong, S., Udalski, A., Gould, A., et al.\ 2007, \apj, 664, 862 

\bibitem[Dunkley et al.(2005)]{dunkley05} 
Dunkley, J., Bucher, M., Ferreira, P.~G., et al.\ 2005, \mnras, 356, 925 

\bibitem[Esplin \& Luhman(2017)]{esplin17} 
Esplin, T.~L., \& Luhman, K.~L.\ 2017, \aj, 154, 134 

\bibitem[Fryer et al.(2012)]{fryer12} 
Fryer, C.~L., Belczynski, K., Wiktorowicz, G., et al.\ 2012, \apj, 749, 91 

\bibitem[Gould(1992)]{gould92}
Gould, A.\ 1992, \apj, 392, 442

\bibitem[Gould(1994)]{gould94} 
Gould, A.\ 1994, \apjl, 421, L75 

\bibitem[Gould(1997)]{gould97}
Gould, A.\ 1997, \apj, 480, 188

\bibitem[Gould et al.(2010a)]{gould10a} 
Gould, A., Dong, S., Bennett, D.~P., et al.\ 2010a, \apj, 710, 1800 

\bibitem[Gould et al.(2010b)]{gould10b} 
Gould, A., Dong, S., Gaudi, B.~S., et al.\ 2010b, \apj, 720, 1073 

\bibitem[Gould et al.(2006)]{gould06} 
Gould, A., Udalski, A., An, D., et al.\ 2006, \apjl, 644, L37 

\bibitem[Gould et al.(2009)]{gould09} 
Gould, A., Udalski, A., Monard, B., et al.\ 2009, \apjl, 698, L147 

\bibitem[Gould \& Yee(2012)]{gould12} 
Gould, A., \& Yee, J.~C.\ 2012, \apjl, 755, L17 

\bibitem[Gould \& Yee(2013)]{gould13}
Gould, A., \& Yee, J. C.\ 2013, \apj, 764, 107

\bibitem[Griest \& Safizadeh (1998)]{griest98}
Griest, K., \& Safizadeh, N.\ 1998, \apj, 500, 37

\bibitem[Han et al.(2016)]{han16} 
Han, C., Udalski, A., Gould, A., et al.\ 2016, \apj, 828, 53 

\bibitem[Han et al.(2017)]{han17} 
Han, C., Udalski, A., Gould, A., et al.\ 2017, \apj, 834, 82 

\bibitem[Juri{\'c} \& Tremaine(2008)]{juric08} 
Juri{\'c}, M., \& Tremaine, S.\ 2008, \apj, 686, 603

\bibitem[Kervella et al.(2004)]{kervella04}
Kervella, P., Bersier, D., Mourard, D., et al.\ 2004, A\&A, 428, 587

\bibitem[Kim et al.(2016)]{kim16} 
Kim, S.-L., Lee, C.-U., Park, B.-G., et al.\ 2016, JKAS, 49, 37

\bibitem[Mao(1999)]{mao99} 
Mao, S.\ 1999, \aap, 350, L19 

\bibitem[Minniti et al.(2010)]{minniti10} 
Minniti, D., Lucas, P. W., Emerson, J. P., et al.\ 2010, NewA, 15, 433 

\bibitem[Mr{\'o}z et al.(2017)]{mroz17} 
Mr{\'o}z, P., Udalski, A., Skowron, J., et al.\ 2017, \nat, 548, 183 

\bibitem[Mr{\'o}z et al.(2018)]{mroz18} 
Mr{\'o}z, P., Ryu, Y.-H., Skowron, J., et al.\ 2018, \aj, 155, 121 

\bibitem[Nataf et al.(2013)]{nataf13}
Nataf, D. M., Gould, A., Fouqu\'e, P., et al.\ 2013, \apj, 769, 88

\bibitem[{\"O}zel et al.(2012)]{ozel12} 
{\"O}zel, F., Psaltis, D., Narayan, R., et al.\ 2012, \apj, 757, 55 

\bibitem[Paczynski(1986)]{paczynski86}
Paczy{\'n}ski, B.\ 1986, \apj, 304, 1 

\bibitem[Poleski et al.(2016)]{poleski16} 
Poleski, R., Zhu, W., Christie, G.~W., et al.\ 2016, \apj, 823, 63 

\bibitem[Refsdal(1966)]{refsdal66}
Refsdal, S.\ 1966, \mnras, 134, 315 

\bibitem[Rhie et al.(2000)]{rhie00} 
Rhie, S.~H., Bennett, D.~P., Becker, A.~C., et al.\ 2000, \apj, 533, 378 

\bibitem[Ryu et al.(2018)]{ryu18} 
Ryu, Y.-H., Yee, J.~C., Udalski, A., et al.\ 2018, \aj, 155, 40 

\bibitem[Shin et al.(2017)]{shin17} 
Shin, I.-G., Udalski, A., Yee, J.~C., et al.\ 2017, \aj, 154, 176 

\bibitem[Shvartzvald et al.(2016)]{shvartzvald16} 
Shvartzvald, Y., Li, Z., Udalski, A., et al.\ 2016, \apj, 831, 183 

\bibitem[Shvartzvald et al.(2015)]{shvartzvald15} 
Shvartzvald, Y., Udalski, A., Gould, A., et al.\ 2015, \apj, 814, 111 

\bibitem[Shvartzvald et al.(2017)]{shvartzvald17} 
Shvartzvald, Y., Yee, J.~C., Calchi Novati, S., et al.\ 2017, \apjl, 840, L3 

\bibitem[Skowron et al.(2016)]{skowron16} 
Skowron, J., Udalski, A., Koz{\l}owski, S., et al.\ 2016, Acta Astron., 66, 1 

\bibitem[Smith et al.(2002)]{smith02} 
Smith, M.~C., Mao, S., Wo{\'z}niak, P., et al.\ 2002, \mnras, 336, 670 

\bibitem[Street et al.(2016)]{street16} 
Street, R.~A., Udalski, A., Calchi Novati, S., et al.\ 2016, \apj, 819, 93 

\bibitem[Sumi et al.(2011)]{sumi11} 
Sumi, T., Kamiya, K., Bennett, D.~P., et al.\ 2011, \nat, 473, 349 

\bibitem[Udalski (2003)]{udalski03} 
Udalski, A.\ 2003, Acta Astron., 53, 291

\bibitem[Udalski et al.(1994)]{udalski94}  
Udalski, A., Szyma\'nski, M., Kaluzny, J., et al.\ 1994, Acta Astron., 44, 227

\bibitem[Udalski et al.(2015a)]{udalski15a}     
Udalski, A., Szyma{\'n}ski, M.~K., \& Szyma{\'n}ski, G.\ 2015, Acta Astron., 65, 1 

\bibitem[Udalski et al.(2015b)]{udalski15b} 
Udalski, A., Yee, J.~C., Gould, A., et al.\ 2015b, \apj, 799, 237 

\bibitem[Wang et al.(2017)]{wang17} 
Wang, T., Zhu, W., Mao, S., et al.\ 2017, \apj, 845, 129 

\bibitem[Wozniak (2000)]{wozniak00}  
Wozniak, P. R.\ 2000, Acta Astron., 50, 421

\bibitem[Whitworth et al.(2007)]{whitworth07}
Whitworth A., Bate M. R., Nordlund \r{A}., Reipurth B., Zinnecker H., 2007,
in Reipurth B., Jewitt D., Keil K., eds, Protostars and Planets V. Tucson,
University of Arizona Press, p. 459

\bibitem[Yee et al.(2015b)]{yee15b} 
Yee, J.~C., Gould, A., Beichman, C., et al.\ 2015b, \apj, 810, 155

\bibitem[Yee et al.(2015a)]{yee15a}  
Yee, J.~C., Udalski, A., Calchi Novati, S., et al.\ 2015a, \apj, 802, 76 

\bibitem[Yee et al.(2009)]{yee09} 
Yee, J.~C., Udalski, A., Sumi, T., et al.\ 2009, \apj, 703, 2082 

\bibitem[Yoo et al. (2004)]{yoo04}
Yoo, Jaiyul, DePoy, D. L., Gal-Yam, A., et al.\ 2004, \apj, 603, 139

\bibitem[Zhu et al.(2016)]{zhu16} 
Zhu, W., Calchi Novati, S., Gould, A., et al.\ 2016, \apj, 825, 60 

\bibitem[Zhu et al.(2017)]{zhu17} 
Zhu, W., Udalski, A., Calchi Novati, S., et al.\ 2017, \aj, 154, 210

\bibitem[Zhu et al.(2015)]{zhu15} 
Zhu, W., Udalski, A., Gould, A., et al.\ 2015, \apj, 805, 8 

\end{thebibliography}
\end{document}